%% file: talk.tex
\documentstyle[sprocl,psfig,epsfig]{article}

\def\ra{\rightarrow}

\def\bea{\begin{eqnarray}}
\def\eea{\end{eqnarray}}

\newcommand{\lsim}{\raisebox{-0.13cm}{~\shortstack{$<$ \\[-0.07cm] $\sim$}}~}
\newcommand{\gsim}{\raisebox{-0.13cm}{~\shortstack{$>$ \\[-0.07cm] $\sim$}}~}

\newcommand{\s}{\\ \vspace*{-3.5mm} }
\newcommand{\nn}{\noindent}

\newcommand{\tb}{\mbox{tan}\beta}
\def\beq{\begin{equation}}
\def\eeq{\end{equation}}

\def\beqn{\begin{eqnarray}}
\def\eeqn{\end{eqnarray}}

\begin{document}

\input title.txt

\title{Top Squark Effects on Higgs Boson Production \\ and Decays at 
the LHC}

\author{Abdelhak DJOUADI}

\address{Laboratoire de Physique Math\'ematique et Th\'eorique, \\
Universit\'e Montpellier II, F--34095 Montpellier Cedex 5\\ 
E-mail: djouadi@lpm.univ-montp2.fr} 


\maketitle\abstracts{In the Minimal Supersymmetric Standard Model, one of the
scalar top quarks can be relatively light and its couplings to Higgs bosons
strongly enhanced. I will discuss two consequences of this feature: 1) the 
effects of scalar top loops on the main production mechanism of the lightest 
Higgs boson at the LHC, the gluon--fusion mechanism $gg \ra h$, and on the 
important decay channel into two photons $h \ra \gamma \gamma$; and 2) the 
production of the light Higgs particle in association with top--squark pairs 
at proton colliders, $pp \ra \tilde{t}_1 \tilde{t}_1 h$.}

\section{Physical Set--Up}

One of the main motivations of supersymmetric theories is the fact that they 
provide an elegant way to break the electroweak symmetry and to stabilize the 
huge hierarchy between the GUT and Fermi scales \cite{R1}. The probing of the 
Higgs sector of the Minimal Supersymmetric Standard Model (MSSM) \cite{R2} is 
thus of utmost importance and the search for the neutral $h,H$ and $A$ and 
charged Higgs particles  $H^\pm$ of the MSSM is therefore one of the main 
entries in the present and future high--energy colliders agendas. 

In the theoretically well motivated models, such as the mSUGRA scenario,
the MSSM Higgs sector is in the so called decoupling regime \cite{decoup} 
for most of the SUSY parameter space allowed by present data  constraints
\cite{data}: the heavy CP--even, the CP--odd  and the charged Higgs bosons are 
rather heavy and almost degenerate in mass, while the lightest neutral CP--even
Higgs  particle reaches its maximal allowed mass value $ M_h \lsim $ 80--130 
GeV \cite{mh} depending on the SUSY parameters. In this scenario, the 
$h$ boson  has almost the same properties as the SM Higgs boson and would be 
the sole Higgs particle accessible at the next generation of colliders. 

At the LHC, the most promising channel \cite{LHCex,LHCth} for detecting the
lightest $h$ boson is the rare decay into two photons, $h \rightarrow \gamma 
\gamma$ \cite{gamma}, with the Higgs particle dominantly produced  via the 
top quark loop  mediated gluon--gluon fusion mechanism \cite{R7}, $gg  \ra h$.
In the decoupling regime, the two LHC collaborations expect to detect the 
narrow $\gamma \gamma$ peak in the entire Higgs mass range, 80 $\lsim M_h
\lsim 130$ GeV, with an integrated luminosity $\int {\cal L} \sim 300$ 
fb$^{-1}$ corresponding to three years of LHC running \cite{LHCex}. 

Two other channels can be used to detect the $h$ boson in this mass 
range \cite{LHCth}:  the production in association with a $W$ boson  or
in association with top quark pairs, $pp \ra hW$ and $pp \ra \bar{t}t h$, 
with the $h$ boson decaying into 2 photons and the $t$ quarks into $b$ quarks 
and $W$ bosons.  Although the cross sections are smaller compared to the $gg 
\ra h$ case, the background cross sections are also  small if one requires 
a lepton from the decaying $W$ bosons as an additional  tag, leading to a 
significant signal. Furthermore, $\sigma( pp \ra \bar{t}th)$ is directly 
proportional to the top--Higgs Yukawa coupling,  the largest electroweak 
coupling in the SM; this process would therefore allow the measurement of 
this parameter, and the experimental test of the fundamental prediction that 
the Higgs couplings to SM particles are proportional to the particle masses. 

In this talk, I will discuss the effects of SUSY 
particles in the production of the lightest $h$ boson at the LHC: first 
the contributions of light stops in the gluon-fusion mechanism $gg \ra h$
and the $h \ra \gamma \gamma$ decay \cite{ggh} which can significantly alter 
the production rate, and then the 
associated production of the $h$ boson with 
$\tilde{t} $--squark pairs, $pp \ra \tilde{t}_1 \tilde{t}_1 h$ \cite{hstop},
for which the cross section can be rather large, 
exceeding the one for the SM--like process $pp \ra t\bar{t}h$. 

One of the most important ingredients of these discussions is that stops can 
alter significantly the phenomenology of the MSSM Higgs bosons. The reason 
is two--fold: $(i)$ the current eigenstates, $\tilde{t}_L$ 
and $\tilde{t}_R$, mix to give the mass eigenstates $\tilde{t}_1$ and 
$\tilde{t}_2$; the mixing angle $\theta_{\tilde{t}}$ is proportional to 
$m_t \tilde{A}_t$ [$\tilde{A}=A_t - \mu/\tan \beta$ with $A_t$ is the 
stop trilinear coupling, $\mu$ the higgsino mass parameter and $\tb$ 
the ratio of the $vev$'s of the two Higgs doublets], and can be very
large, leading to a scalar top squark $\tilde{t}_1$ much  lighter than the
$t$--quark and all other scalar quarks [note that the mixing in the sbottom 
sector can be also sizeable for large values of $\tb,\mu$ and $A$; however 
this will not be discussed here and the $\tilde{b}$ mixing will be set to 
zero]; $(ii)$ the couplings of the top 
squarks to the neutral Higgs bosons in the decoupling regime 
$$
g_{h \tilde{t}_1 \tilde{t}_1 } \propto - \cos 2\beta \left[ 
\frac{1}{2} \cos^2 \theta_{\tilde{t}} - \frac{2}{3} s^2_W \cos 2
\theta_{\tilde{t}} \right] 
- \frac{m_t^2}{M_Z^2} + \frac{1}{2} \sin 2\theta_{\tilde{t}} 
\frac{m_t \tilde{A}_t } {M_Z^2} \hspace*{0.5cm} (1) 
$$ 
involve components which are proportional to $\tilde{A}_t$ and for large 
values of this parameter, the $g_{h\tilde{t}_1 \tilde{t}_1}$ coupling can 
be strongly enhanced.

\section{Higgs production in the gluon fusion mechanism}

In the SM, the Higgs--gluon--gluon vertex is mediated by heavy 
[mainly top and to a lesser extent bottom] quark loops, while the rare 
decay into two photons is mediated by $W$--boson and heavy fermion 
loops, with the $W$--boson contribution being largely dominating. In the 
MSSM however, additional contributions are provided by SUSY particles: 
$\tilde{q}$ loops in the case of the $hgg$ vertex, and $H^\pm, \tilde{f}$
and $\chi^\pm$ loops in the case of the $h \ra \gamma \gamma$ decay. 
In Ref.~\cite{gamma}, it has been shown that only the contributions of 
relatively light $\tilde{t}$ squarks [and to a lesser extent $\chi_1^\pm$ 
for masses close to 100 GeV, which could contribute at the 10\% level] can 
alter significantly the loop induced $hgg$ and $h \gamma \gamma$ vertices. 
The $\gamma \gamma$ and $gg$ decay widths [the $gg$ cross section is 
proportional to the partial width] of the $h$ boson, including the SUSY 
loops and the QCD corrections, are evaluated numerically with the help of 
the program HDECAY \cite{HDECAY}. \s

Figs.~1 and 2 show, as a function of $\tilde{A}_t$ for $\tb=2.5$, the 
deviations from their SM values of the partial decay widths of the $h$ boson 
into two photons and two gluons as well as their product which gives the 
cross section times branching ratio $\sigma(gg \ra h \ra \gamma \gamma)$. The 
quantities $R$ are defined as the partial widths including, over the ones 
without [equivalent to SM in the decoupling limit] the SUSY loop contributions 
$R=\Gamma_{\rm MSSM}/\Gamma_{\rm SM}$. \s

\begin{figure}[htb]
\vspace*{-.5cm}
\mbox{\psfig{figure=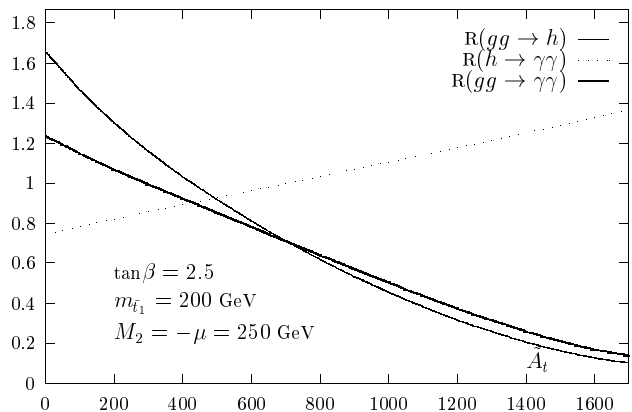,width=12cm}}
\vspace*{-11.2cm}
\caption[]{SUSY loop effects on R$(gg \ra h)$, R$(h \ra \gamma \gamma)$ and
R$(gg \ra \gamma \gamma)$ as a function of $\tilde{A}_{t}$ for $\tb=2.5$ and 
$m_{\tilde{t}_1}=200$ GeV, $M_2=-\mu=250$ GeV.}
\end{figure}
\begin{figure}[htb]
\vspace*{-.5cm}
\mbox{\psfig{figure=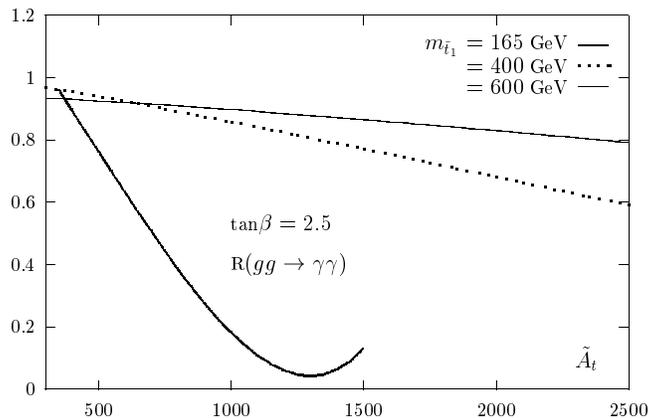,width=12cm}}
\vspace*{-11.2cm}
\caption[]{SUSY loop effects on R$(gg \ra \gamma \gamma)$ as a function of 
$\tilde{A}_{t}$ for $\tb=2.5$ and $m_{\tilde{t}_1}=165,400$ and 600 GeV with 
$M_2=-\mu=500$ GeV for $m_{\tilde{t}_1}\geq 400$ GeV.} 
\vspace*{-.5cm}
\end{figure}

In Fig.~1, the stop mass is set to $m_{\tilde{t}_1}=200$ GeV. For small values 
of $\tilde{A}_t$ there is no mixing in the stop sector and the dominant 
component of the $h\tilde{t} \tilde{t}$ couplings eq.~(1) is $\sim m_t^2/M_Z^2$.
In this case, the $t$ and $\tilde{t}$ contributions interfere constructively 
in the $hgg$ and $h\gamma \gamma$ amplitudes, which leads to an enhancement of 
the $h \ra gg$ decay width and a reduction of the $h \ra \gamma \gamma$ decay 
width [that is dominated by the $W$ amplitude which interferes destructively 
with the $t$ and $\tilde{t}$ amplitudes]. The product R($gg \ra \gamma \gamma$)
in the MSSM is then enhanced by a factor $\sim 1.2$ in this case. 
With increasing $\tilde{A}_{t}$, the two components of $g_{h\tilde{t}_1 
\tilde{t}_1}$ interfere destructively and partly cancel each other, resulting 
in a rather small stop contribution. For larger values of $\tilde{A}_{t}$, the 
last component of $g_{h\tilde{t}_1 \tilde{t}_1}$ becomes the most important 
one, and the $\tilde{t}_1$ loop contribution interferes destructively with the 
$t$--loop. This leads to an enhancement of R$(h \ra \gamma \gamma)$ and a 
reduction of R$(gg \ra h)$; however, the reduction of the latter is much 
stronger than the enhancement of the former and the product R($gg \ra \gamma 
\gamma$) decreases with increasing $\tilde{A}_t$. For $\tilde{A}_t$ values of 
about 1.5 TeV, the signal for $gg\ra h \ra \gamma \gamma$ in the MSSM is 
smaller by a factor of $\sim 5$ compared to the SM case. 

Fig.~2 shows the deviation  R$(gg \ra \gamma \gamma)$ with the same parameters
as in Fig.~1 but with different $\tilde{t}_1$ masses, $m_{\tilde{t}_1}=
165,400$ and 600 GeV. For larger masses, the top squark contribution 
$\propto 1/ m_{\tilde{t}_1}^2$, will be smaller than in the previous case. 
For $m_{\tilde{t}_1}\simeq 400$ GeV, the effect is less striking compared to 
the case of $m_{\tilde{t}_1}=200$ GeV, since here $\sigma(gg \ra h) \times 
{\rm BR} (h \ra \gamma \gamma$) drops by less than a factor of 2, even for 
extreme values of $\tilde{A}_t \sim 2.5$ TeV. In contrast, if the stop mass 
is reduced to $m_{\tilde{t}_1} \simeq 165$ GeV, the drop in  R$(gg \ra \gamma 
\gamma)$ will be even more important: for  $\tilde{A}_t \sim 1.5$ TeV, the $gg 
\ra \gamma \gamma$ rate including stop loops 
is an order of magnitude smaller than in the SM. For $\tilde{A}_t \sim 1.3$ 
TeV, the $\tilde{t}$ amplitude almost cancels completely the $t/b$ quark 
amplitudes. 

One should recall that $M_h$ varies with $\tilde{A}_t$, and no constraint on 
$M_h$ has been set in Figs.~1--2. Requiring $M_h \gsim 90$ GeV, the 
lower range $\tilde{A}_t \lsim 350$ GeV and the upper ranges $\tilde{A}_t 
\gsim 1.5 (2.3)$ TeV for $m_{\tilde{t}_1}=200 (400)$ GeV for instance, are 
ruled out. This means that the scenario where R$(gg \ra \gamma \gamma) >1$, 
which occurs only for $\tilde{A}_t \lsim 300$ GeV for $m_{\tilde{t}_1}=200$ GeV
is ruled out for $M_h \gsim 90$ GeV. [Note also that despite of the large
$(\tilde{t},\tilde{b}$) mass splitting for large $\tilde{A}_t$ values, the 
contributions of the isodoublet to electroweak observables \cite{drho} stay 
below the 
experimentally acceptable level]. Therefore, the rate for the $gg \ra \gamma 
\gamma$ process in the MSSM will always be smaller than in the SM case, making
more delicate the search for the $h$ boson at the LHC with this process. 

\section{Higgs production in association with light stops} 

If one of the stop squarks is light and its coupling to the $h$ boson is 
enhanced, an additional process might provide a new important source for 
Higgs particles: the associated production with $\tilde{t}$ states, 
\beq
pp \ra gg + q \bar{q} \ra \tilde{t}_1 \tilde{t}_1 h
\eeq 
At lowest order, i.e. at ${\cal O}( G_F \alpha_s^2)$, the process is initiated 
by 12 Feynman diagrams. Due to the larger gluon luminosity at high energies, 
the contribution of the 10 $gg$--fusion diagrams is much larger  than the 
contribution of the 2 $q\bar{q}$ annihilation diagrams at the LHC. 

In Fig.~3, the $pp \ra \tilde{t}_1 \tilde{t}_1h$ cross section [in pb] is 
displayed as a function of $m_{\tilde{t}_1}$ for $\tb=2$, in the case of 
no--mixing [$A_t=200, \mu=400$ GeV], moderate mixing [$A_t=500$ and $\mu=100$ 
GeV] and large mixing [$A_t=1.5$ TeV and $\mu=100$ GeV]. We have used 
$m_{\tilde{t}_L}= m_{\tilde{t}_R} \equiv m_{\tilde{q}}$ as is approximately 
the case in GUT  scenarios. Note for comparison, that the cross section for 
the standard--like $pp \ra \bar{t} t h$ process is of the order of 0.6 pb 
for $M_h \simeq 100$ GeV \cite{LHCth}; $m_t=175$ GeV, and the CTEQ4 
parameterizations of the structure functions \cite{CTEQ} are chosen
for illustration. 

If there is no mixing in the stop sector, $\tilde{t}_1$  and
$\tilde{t}_2$ have the same mass and approximately the same couplings to the 
$h$ boson since the $m_t^2/M_Z^2$ components are dominant. The cross  section,
which should be then multiplied by a factor of two to take into account  both
squarks, is comparable to the $\sigma(pp \ra t\bar{t} h)$ in the  low
mass range $m_{\tilde{t}}\lsim  200$ GeV. [If the  
$\tilde{t}$ masses are related to the masses of the light quark partners,  
$m_{\tilde{q}}$, the range for which the cross section is rather large is
therefore ruled out by the experimental constraints on $m_{\tilde{q}}$ 
\cite{data}.] For intermediate values of $\tilde{A}_t$ the two components of 
the  $h \tilde{t}_1 \tilde{t}_1$ coupling interfere destructively and partly 
cancel each other, resulting in a rather small cross section, unless
$m_{\tilde{t}_1} \sim {\cal O}(100)$ GeV. 
In the large mixing case $\tilde{A}_t \sim 1.5$ TeV $\sigma(pp \ra \tilde{t}_1 
\tilde{t}_1 h)$ can be very large. It is above the rate for the standard 
process $pp \ra \bar{t}th$ for values of $m_{\tilde{t}_1}$ smaller than 220 GeV.
If  $\tilde{t}_1$ is lighter than the top quark, the $\tilde{t}_1 \tilde{t}_1 
h$ cross section significantly exceeds the one for $\bar{t}th$ final states.
For instance, for $m_{\tilde{t}_1}=140$ GeV corresponding to $M_h\sim 76$ GeV, 
$\sigma( pp \ra \tilde{t}_1 \tilde{t}_1 h)$ is an order of magnitude larger 
than $\sigma(pp \ra t\bar{t}h)$. 

\begin{figure}[htb]
\begin{center} 
\vspace*{-.6cm}
\mbox{\psfig{figure=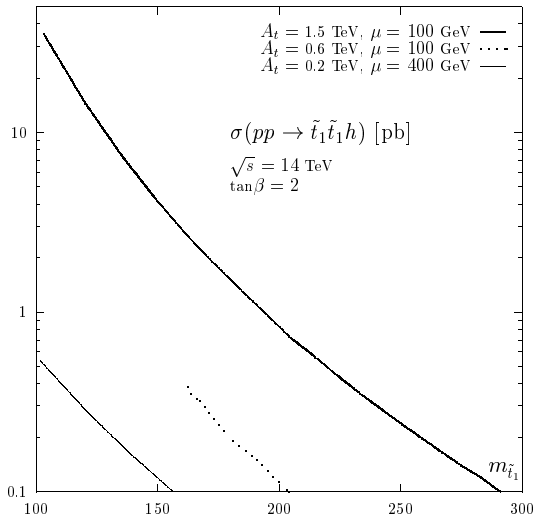,width=10cm}}
\vspace*{-6.8cm}
\caption[]{The production cross section $\sigma (pp \ra \tilde{t}_1 
\tilde{t}_1 h$) [in pb] as a function of the $\tilde{t}_1$ mass and three 
sets of $A_t$ and $\mu$ values and $\tb$ is fixed to $\tb=2$.}
\end{center}
\vspace*{-6.8mm}
\end{figure}

In Fig.~4, we fix $m_{\tilde{t}_1} =165$ GeV 
$ \sim m_{t}^{\rm \overline{MS}}$ and display the $pp \ra 
\tilde{t}_1 \tilde{t}_1h$ cross section as a function of $\tilde{A}_t$. 
For comparison, the $*$ and $\bullet$ give the standard--like  $pp \ra 
\bar{t}th$ cross section for $M_h=100$ GeV and $\tb=2$ and 30, respectively. 
For $\tb=30$ the cross section is somewhat smaller than for $\tb=2$, a mere 
consequence of the increase of the $h$ boson mass with $\tb$ \cite{mh}.  As 
can be seen again, the production cross  section is substantial for the 
no--mixing case, rather small for intermediate mixing [becoming negligible for 
$\tilde{A}_t$ values between 200 and 400 GeV], and then becomes very large 
exceeding the reference cross section for values of $\tilde{A}_t$  above $\sim 
1$ TeV. For instance, for the inputs of Fig.~3, $\sigma(pp \ra
\tilde{t}_1 \tilde{t}_1 h)$ exceeds $\sigma(pp \ra t\bar{t}h)$ in the SM for the
same Higgs boson mass when $\tilde{A}_t \gsim 1(1.05)$ TeV for $\tb=2(30)$. 

\begin{figure}[htb]
\begin{center} 
\vspace*{-.6cm}
\mbox{\psfig{figure=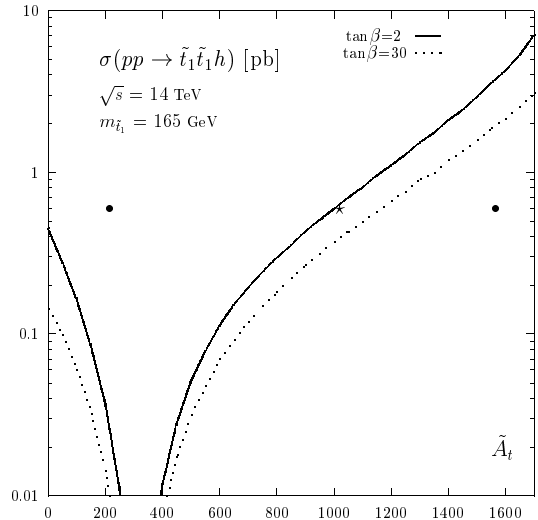,width=10cm}}
\vspace*{-6.8cm}
\caption[]{The production cross section $\sigma (pp \ra \tilde{t}_1 
\tilde{t}_1 h$) [in pb] as a function as a function of $\tilde{A}_t$ for
 fixed $m_{\tilde{t}_1}=165$ GeV and for $\tb=2,30$.}
\end{center}
\vspace*{-6.8mm}
\end{figure}

For the signal, in most of the parameter space, the stop decay is $\tilde{t}_1 
\ra b\chi^+$ if  $m_{\tilde{t}_1} <m_t+m_{\chi_1^0}$  where $\chi_1^0$ is the 
LSP, or $\tilde{t}_1 \ra t \chi_1^0$ in the opposite case. 
In the interesting region where  $\sigma(pp \ra \tilde{t}_1  \tilde{t}_1 h)$ 
is large, i.e. for light $\tilde{t}_1$, the decay  $\tilde{t}_1 \ra b \chi^+$ 
is dominant, and the $\chi_1^+$ will mainly decay into $bW^+ + \ {\rm missing 
\ energy}$ leading to $\tilde{t}_1 \ra bW^+$ final states. This is the same 
topology as the decay $t \ra bW^+$, except that in the case of the 
$\tilde{t}$ there is a large amount of missing energy [if sleptons are also 
relatively light, charginos decays will also lead to $l \nu \chi_1^0$ final 
states]. The only difference between the final states generated by the 
$\tilde{t} \tilde{t}h$ and $t\bar{t}h$ processes, will be due to the softer 
energy spectrum of the charged leptons coming from the chargino decay in the 
former case, because of the energy carried by the invisible LSP. 
The Higgs boson can be tagged through its $h \ra \gamma \gamma$ decay mode;
as discussed previously this mode can be substantially enhanced compared to 
the SM case for light top squarks and large $\tilde{A}_t$ values. Therefore, 
$\gamma \gamma$+ charged lepton events can be  more copious than in the SM, 
and the contributions of the $pp \ra \tilde{t} \tilde{t} h$ process to these 
events can render the detection of the $h$ boson much easier than with the 
process $pp \ra t \bar{t}h$ alone. 

\section{Conclusions}
 
I discussed the effects of $\tilde{t}$ squarks on the production  the lightest 
neutral SUSY Higgs boson $h$ at the LHC in the context of the minimal 
supersymmetric extension of the Standard Model. 

If the off--diagonal entries in the $\tilde{t}$  mass matrix are large, the
eigenstate $\tilde{t}_1$ can be rather light and at the same time its 
couplings to the $h$ boson strongly enhanced. The stop loops can then 
make the cross section times branching ratio $\sigma( gg \ra h) \times {\rm 
BR}(h \ra \gamma \gamma)$, which is the main production and detection 
mechanism for Higgs bosons at 
the LHC, much smaller than in the SM, even in the decoupling 
regime where the $h$--boson has SM--like couplings to fermions and 
gauge bosons. [Far from this decoupling limit, the cross section times 
branching ratio is further reduced in general due to the additional 
suppression of the $ht\bar{t}$ and $hWW$ couplings.] This will render
more delicate the search for the $h$ boson at the LHC with this process. 

In this scenario, the process $pp \ra \tilde{t}_1 \tilde{t}_1h$ can be
a copious source of Higgs bosons at the LHC, since the cross section
can exceed the one for the SM--like $t\bar{t}h$ production process 
[for the heavier $H$ and $A$ bosons, the processes are phase space 
suppressed \cite{prepa}]. This
can significantly enhance the potential of the LHC to discover the lightest 
MSSM Higgs boson in the $l^+ \gamma \gamma$ channel. 
As a bonus, this process 
would  allow to measure the $h\tilde{t} \tilde{t}$ coupling, the potentially
largest electroweak coupling in the MSSM, opening thus a window to probe 
directly the trilinear part of the soft--SUSY breaking scalar potential.  \s

\nn {\bf Acknowledgements:} I thank the organizing committee, and in 
particular Joan Sol\`a, for the very nice athmosphere of this fruitful 
workshop. 


\end{document}

%% file: title.txt
\thispagestyle{empty}

\begin{flushright}
PM--98/44 \\
December 1998
\end{flushright}

\vspace*{1cm}

\begin{center}

\Large{\bf Top Squark Effects on Higgs Boson Production \\ and Decays at 
the LHC}

\vspace*{1cm} 

\large{\sc Abdelhak DJOUADI} 

\vspace*{1cm}

Laboratoire de Physique Math\'ematique et Th\'eorique,  \\
Universit\'e Montpellier II, F--34095 Montpellier Cedex 5, France. \\
E-mail: djouadi@lpm.univ-montp2.fr \\

\vspace*{2cm} 

{\bf ABSTRACT}

\bigskip 

\end{center} 

\noindent 
In the Minimal Supersymmetric Standard Model, one of the
scalar top quarks can be relatively light and its couplings to Higgs bosons
strongly enhanced. I will discuss two consequences of this feature: 1) the 
effects of scalar top loops on the main production mechanism of the lightest 
Higgs boson at the LHC, the gluon--fusion mechanism $gg \ra h$, and on the 
important decay channel into two photons $h \ra \gamma \gamma$; and 2) the 
production of the light Higgs particle in association with top--squark pairs 
at proton colliders, $pp \ra \tilde{t}_1 \tilde{t}_1 h$.

\vspace*{3cm}

\begin{center}
Talk given at the Workshop  RADCOR'98, 
Barcelona, Spain, September  1998. 
\end{center}

\setcounter{page}{0}

\newpage  